\documentclass{mn2e}
\title{The role of tidal interactions in star formation}

\author[Richard B. Larson]
       {Richard B. Larson\thanks{E-mail: larson@astro.yale.edu} \\
  Yale Astronomy Department, Box 208101, New Haven, CT 06520-8101, USA}

\date{Submitted to MNRAS 30 August 2001; revised 28 November 2001}

\pagerange{\pageref{firstpage}--\pageref{lastpage}}
\pubyear{2002}
\begin{document}
\maketitle
\label{firstpage}

\begin{abstract}

   Nearly all of the initial angular momentum of the matter that goes
into each forming star must somehow be removed or redistributed during
the formation process.  The possible transport mechanisms and the
possible fates of the excess angular momentum are discussed, and it is
argued that transport processes in disks are probably not sufficient
by themselves to solve the angular momentum problem, while tidal
interactions with other stars in forming binary or multiple systems
are likely to be of very general importance in redistributing angular
momentum during the star formation process.  Most, if not all, stars
probably form in binary or multiple systems, and tidal torques in
these systems can transfer much of the angular momentum from the gas
around each forming star to the orbital motions of the companion stars.
Tidally generated waves in circumstellar disks may contribute to the
overall redistribution of angular momentum.  Stars may gain much of
their mass by tidally triggered bursts of rapid accretion, and these
bursts could account for some of the most energetic phenomena of the
earliest stages of stellar evolution, such as jet-like outflows.  If
tidal interactions are indeed of general importance, planet-forming
disks may often have a more chaotic and violent early evolution
than in standard models, and shock heating events may be common.
Interactions in a hierarchy of subgroups may play a role in building up
massive stars in clusters and in determining the form of the upper IMF.
Many of the processes discussed here have analogs on galactic scales,
and there may be similarities between the formation of massive stars
by interaction-driven accretion processes in clusters and the buildup
of massive black holes in galactic nuclei.

\end{abstract}


\section{INTRODUCTION}

   A long-standing problem in the theory of star formation is the
`angular momentum problem', or the problem of understanding how most of
the initial angular momentum is removed from the material that condenses
into each star.  Because the sizes of stars are tiny compared with
the sizes of the cloud cores in which they form, individual stars can
contain only a tiny fraction of the angular momentum typically found
in these cores; therefore nearly all of the angular momentum of the gas
that goes into each star must somehow be removed or redistributed during
the star formation process (Mestel \& Spitzer 1956; Mestel 1965;
Spitzer 1968; Bodenheimer 1995.)  What mechanisms might account for
this efficient removal of angular momentum, and where does the excess
angular momentum go?  Understanding the fate of the excess angular
momentum is an important aspect of the problem, because this can put
strong constraints on the mechanisms involved; however, the fate of
the angular momentum that is removed has received relatively little
attention in most discussions of the problem.

   One possibility that has been much studied is that viscous torques
in a protostellar accretion disk carry away most of the angular momentum
of the gas that goes into each star and deposit it in the outer part of
the disk (Shu, Adams \& Lizano 1987; Papaloizou \& Lin 1995; Stone et
al.\ 2000; Stahler 2000).  A second possibility that has long been
recognized is that most of the angular momentum of a collapsing cloud
core goes into the orbital motions of the stars in a binary or multiple
system (Mestel \& Spitzer 1956; Larson 1972; Mouschovias 1977); this
is plausible because most stars do indeed form in binary or multiple
systems, and because the typical angular momentum of a star-forming
cloud core is comparable to that of a wide binary (Bodenheimer 1995;
Larson 1997, 2001).  If most of the excess angular momentum of the gas
going into each star ends up in the orbital motions of the companion
stars in a binary or multiple system, protostellar interactions must
then play a central role in the star formation process, and tidal
torques must transfer most of the excess angular momentum from the gas
that forms each star to the orbital motions of its companions.

   The current status of the angular momentum problem and the
possible mechanisms that might redistribute angular momentum during
star formation are reviewed in Section~2, and it is argued there that
internal transport processes in disks are probably not sufficient by
themselves to solve the problem, but that tidal interactions with
companion stars in forming binary or multiple systems can be of quite
general importance in redistributing angular momentum during the star
formation process.  The dynamical evolution of circumstellar disks is
discussed further in Section~3, and it is suggested that even if no
companion is present initially, one or more companions may eventually
be formed by the fragmentation of such a disk.  Any companions thus
formed, including massive planets, can then play an important role in
the further evolution of the system through their tidal effects on the
remaining disk.  The formation of companions may then be a very general
feature of star formation, and it may be how nature solves the angular
momentum problem.

   If protostellar interactions and tidal torques play a major role in
star formation, this will have important implications for the earliest
stages of stellar evolution as well as for the properties of stellar
and planetary systems.  Section~4 discusses some of the implications of
tidal interactions for early protostellar evolution, planet formation,
the formation of massive stars, and the origin of the stellar IMF;
also discussed is the possibility that similar effects may play a role
in the growth of black holes in galactic nuclei. The main conclusions
of the paper are summarized in Section~5.

\section{THE ANGULAR MOMENTUM PROBLEM}

   Early discussions of the angular momentum problem considered the
angular momentum of a solar mass of material in an interstellar cloud
rotating with the Galaxy, and showed that this is many orders of
magnitude more than can be accommodated in a single star even when
rotating at breakup speed (Mestel 1965; Spitzer 1968).  We have learned
since then that stars form in dense molecular cloud cores that rotate
much more slowly than would be expected on this basis (Goodman et al.\
1993), plausibly because magnetic fields have already removed much of
the initial angular momentum during early stages of cloud contraction
(Mestel 1965; Mouschovias 1977, 1991).  However, magnetic fields are
predicted to decouple from the gas in star-forming cloud cores by
ambipolar diffusion long before stellar densities are reached, and
angular momentum is then approximately conserved during the later stages
of the collapse (Basu \& Mouschovias 1995; Basu 1997; Mouschovias \&
Ciolek 1999).  Observations of collapsing cloud cores confirm that
angular momentum is approximately conserved on scales smaller than
a few hundredths of a parsec (Ohashi et al.\ 1997; Ohashi 1999; Myers,
Evans, \& Ohashi 2000).  An important part of the angular momentum
problem therefore remains unsolved, since the observed typical angular
momentum of the collapsing cloud cores is still three orders of
magnitude larger than the maximum that can be contained in a single star
(Bodenheimer 1995).  This means that only a tiny fraction of the mass
in a star-forming core can end up in a star unless non-magnetic torques
remove or redistribute most of the remaining angular momentum.

   Where does this excess angular momentum go?  In principle, there
are two possibilities: it could go into residual diffuse gas at large
distances from the star, for example gas in an extended disk as in
standard protostellar accretion disk models, or it could go into the
orbital motions of the stars in a binary or multiple system.  In the
first case, the excess angular momentum must somehow be transported
to large radii in the residual gas, for example to the outer part of
an extended circumstellar disk, while in the second case, the excess
angular momentum of the gas forming each star must be transferred to
the orbital motions of its companions.  We consider below the mechanisms
that might be involved in each case.  In principle, a centrifugally
driven disk wind could also remove angular momentum from a protostellar
disk (Ouyed \& Pudritz 1999; K\"onigl \& Pudritz 2000), but completely
self-consistent models based on this hypothesis have not yet been
developed.  In any case, if the wind is launched from a small region
close to the central star, as seems likely, it cannot contribute much
to the solution of the angular momentum problem because the matter in
this region must already have lost most of its initial angular momentum.

\subsection{Transport processes in disks}

   In the `standard model' for isolated star formation (Shu, Adams, \&
Lizano 1987), a central stellar object of very small mass forms first,
and a disk is then built up around it by the infall of gas that has too
much angular momentum to fall directly onto the star.  The star acquires
most of its final mass from the disk via an accretion flow driven by
an assumed viscosity that transfers angular momentum outward and mass
inward; most of the angular momentum of the system then ends up in
the outer part of the disk.  In order to absorb all of this angular
momentum, the disk must expand by a large factor in radius (Bodenheimer
1995; Calvet, Hartmann, \& Strom 2000); for example, if the residual
disk has 25 per cent of the initial mass, it must expand by a factor
of 16 in radius, while if it has 10 per cent of the initial mass, it
must expand by a factor of 100, becoming larger in this case than the
original cloud core.  Since most of the angular momentum resides in
the outer part of the disk, most of the angular momentum must then be
transported out to these very large radii.

   This simple model, however, encounters both observational and
theoretical difficulties.  Observationally, the residual disks seen
around most nascent stars are too small in size and mass to contain a
significant fraction of the angular momentum of a collapsing cloud core
(Beckwith 1999; Mundy, Looney, \& Welch 2000).  On the theoretical
front, despite extensive study, no mechanism has yet been identified
that is clearly capable of transporting angular momentum throughout an
extended protostellar disk (Larson 1989; Adams \& Lin 1993; Papaloizou
\& Lin 1995; Stone et al.\ 2000; Stahler 2000.)  The effect originally
suggested by Shakura and Sunyaev (1973), hydrodynamic turbulence, has
no clear justification, since disks are stable against the spontaneous
development of turbulence.  Even if turbulence is somehow generated in
a disk by an external stirring mechanism, it can only transport angular
momentum relatively inefficiently by a diffusion process, and it is
not clear that this process will actually transport angular momentum
outward, as required.  Numerical simulations show that turbulence is
indeed inefficient as a transport mechanism, and moreover that it tends
to transport angular momentum {\it inward}\/, not outward as required
in standard disk models (Stone et al.\ 2000; Quataert \& Chiang 2000).
Therefore hydrodynamic turbulence does not appear promising as a
mechanism for transporting angular momentum outward in disks.

   More recently, much attention has been given to the magnetorotational
or `Balbus-Hawley' instability as a transport mechanism that can be
much more effective than hydrodynamic turbulence if there is sufficient
ionization present (Stone et al.\ 2000).  The outermost part of a
protostellar disk may be kept sufficiently ionized by ambient cosmic
rays, and the part inside 0.1~au may be kept ionized by radiation from
the central star, but this still leaves an extended `dead zone' at
intermediate radii where ionization is negligible and this mechanism
cannot operate, except perhaps in a surface layer (Gammie 1996).  Even
if the entire region inside 1~au could be kept sufficiently ionized by
self-sustaining MHD turbulence, as suggested by Terquem (2001), this
would still leave much of the mass of the disk in the dead zone.  In
any case, this mechanism yields accretion rates that are only marginally
sufficient to be important for star formation (Stahler 2000), so it
probably does not contribute much to the solution of the angular
momentum problem.  It might still be of some importance, however, for
the final stages of protostellar accretion after most of the stellar
mass has been accreted, and it might help to explain the evidence for
continuing accretion of matter at a low rate by many T~Tauri stars.

   If viscous torques do not transport angular momentum effectively,
infalling matter may accumulate in a disk until its self-gravity becomes
important, and gravitational effects may then dominate the subsequent
evolution of the system (Larson 1984, 1989; Adams \& Lin 1993;
Bodenheimer 1995; Stone et al.\ 2000; Stahler 2000; Gammie 2001).
One possibility suggested by these authors is that weak gravitational
instabilities in a marginally stable disk generate transient spiral
density fluctuations whose gravitational torques transfer angular
momentum outward and drive an accretion flow.  However, most observed
protostellar disks have masses that are at least an order of magnitude
too small for such self-gravitational effects to be important (Beckwith
1999; Mundy et al.\ 2000).  It is not clear, in any case, whether a
marginally stable state can be created and sustained, since this
requires fine tuning: the disk must be assembled carefully in a way that
brings it just to, but not beyond, the threshold of stability without
significant disturbances, since otherwise it may break up into clumps.
Such an idealized formation processes may not often occur in reality
because real star-forming cores are turbulent and irregular and show
structure on all resolvable scales that probably typically forms
binary or multiple systems (Mundy, Looney, \& Welch 2000, 2001).  Any
circumstellar disk that forms in such a complex environment will be
subject to continuing strong perturbations, and if it acquires enough
mass, it may fragment into clumps rather than remain in a marginally
stable state.

   Even if one or more of the transport mechanisms mentioned above
does operate, the timescale for angular momentum transport becomes
unacceptably long if the disk expands to the large radii required by
standard accretion disk models.  For example, if the disk expands to a
radius larger than 1000~au, the transport timescale becomes longer than
$10^5$~yr even in the most optimistic case where the alpha parameter
is of order unity (Larson 1989).  More realistic values would yield
timescales that are much longer than this, and longer than any observed
timescale associated with star formation.  Therefore the transport of
angular momentum to large radii by internal processes in protostellar
disks does not appear promising as a solution to the angular momentum
problem.  If internal transport mechanisms are not effective and
infalling matter continues to accumulate in a disk, the disk may then
become strongly gravitationally unstable and may fragment into one or
more companion objects.  Once this has happened, interactions between
the newly formed companions and the residual disk will dominate the
subsequent evolution of the system, and additional transport processes
such as tidally generated waves in disks can begin to play a role, as
will be discussed further in Section~3.2.

\subsection{Effects of interactions and tidal torques}

   If transport processes in disks are not sufficient by themselves to
solve the angular momentum problem, the alternative possibility is that
most of the angular momentum in a collapsing cloud core goes into the
orbital motions of the stars in a binary or multiple system (Mouschovias
1977; Spitzer 1978; Bodenheimer 1995; Larson 2001).  Even in the context
of standard accretion disk models, a binary companion may play an
important role by limiting the growth in radius of the disk and thus
preventing the accretion time from becoming too long (Calvet, Hartmann,
\& Strom 2000); in this case most of the angular momentum of the system
must go into the orbital motion of the companion.  Since gravity is the
only force capable of significantly altering stellar orbital motions,
gravity must be the force that ultimately transfers the excess angular
momentum from the gas around each forming star to the orbital motions
of the companions in a binary or multiple system.  The tidal torques
exerted by the companions on the gas orbiting around each forming star
must then play a fundamental role in the dynamics of the system and in
the evolution of any remaining circumstellar disk.

   Both analytical theory (Ostriker 1994) and numerical simulations
(Heller 1993, 1995) show that tidal interactions typically remove
angular momentum from circumstellar disks, regardless of the orbit of
the perturber.  This effect can be substantial; for example, a single
flyby can reduce the angular momentum of a disk by as much as 10 per
cent (Ostriker 1994).  Thus, multiple encounters in a forming cluster
of stars could in principle extract much of the angular momentum from
a circumstellar disk and drive protostellar accretion (Larson 1982,
1990b), but the frequency of close passages is probably too small for
purely random encounters to be very important, even in dense clusters
like the Orion Nebula Cluster (Scally \& Clarke 2001).  The importance
of interactions is however greatly increased if most stars, even in
clusters, actually form in close proximity to other stars in binary or
multiple systems (Larson 1995; Simon 1997).  Interaction with a binary
companion can be much more important than interactions with passing
stars in a cluster because a binary companion can continually strongly
perturb a circumstellar disk (Lubow \& Artymowicz 2000).  Numerical
simulations of forming binary systems show that tidal torques can indeed
be very effective in transferring angular momentum from circumstellar
disks to stellar orbits and thus in driving accretion onto the forming
stars (Bate 2000; Nelson 2000).  More generally, tidal torques may be
important in driving accretion flows in many types of binary systems
containing circumstellar disks (Matsuda et al.\ 2000; Menou 2000).

   The simulations mentioned above have all assumed nearly circular
orbits for the stars, but tidal effects may be even more important in
the more typical case of eccentric binaries, since the two stars then
make repeated close passages.  In an eccentric binary with a typical
period of the order of 100 years, 1000 or more close passages may
occur during the time when most of the stellar mass is being accreted,
and much of the mass may then be acquired during episodes of enhanced
accretion triggered by the resulting strong tidal disturbances, as
illustrated  by the numerical simulations of Bonnell \& Bastien (1992).
Tidally induced bursts of rapid accretion could plausibly account for
some of the most energetic phenomena characterizing the very earliest
stages of stellar evolution, such as FU Orionis flareups (Bonnell
\& Bastien 1992) and Herbig-Haro jets (Reipurth 2000), as will be
discussed further in Section~4.1.

   Given the high frequency of binary and multiple systems and the
possibility that {\it all} stars form in such systems (Heintz 1969; Abt
1983; Larson 1972, 2001), it seems very likely that most of the angular
momentum in star-forming cloud cores ends up in the orbital motions
of the stars in these systems.  The angular momentum of a typical
collapsing cloud core is comparable to that of a wide binary (Simon et
al.\ 1995; Bodenheimer 1995), and this suggests that wide binaries may
form directly by the collapse of such cores.  Close binaries can only
form in the same way if there is a substantial population of cloud
cores with angular momenta much smaller than the detectable limits,
but simulations of turbulent velocity fields (Burkert \& Bodenheimer
2000) do not support this possibility and do not predict enough cores
with very small angular momenta to account for the observed broad
distribution of  binary periods.  However, the dispersion in binary
properties can plausibly be accounted for if most binaries actually
form in more complex systems where interactions with dense surrounding
matter, for example other forming stars, can efficiently remove energy
and angular momentum from the orbits of the forming binaries (Larson
1997, 2001; Heacox 1998, 2000).  For example, if most stars actually
form in multiple systems that disintegrate into a combination of
binaries and single stars, the chaotic dynamics of these systems can
create binaries with a wide range of properties, and this can account
for at least part of the broad spread in binary periods (Sterzik \&
Durisen 1998, 1999; Larson 2001).  The formation of multiple systems may
be a quite general outcome of star formation, since this is suggested
by many simulations of the collapse and fragmentation of cloud cores
(e.g., Larson 1978; Burkert, Bate, \& Bodenheimer 1997; Bodenheimer et
al.\ 2000; Boss 2001a; Whitworth 2001; Klein, Fisher, \& McKee 2001).

   Theory and observation are thus both consistent with the hypothesis
that most of the angular momentum of collapsing cloud cores goes into
the orbital motions of the stars in binary or multiple systems.
Protostellar interactions and tidal torques must then play a central
role in redistributing angular momentum during star formation and in
allowing stars to accrete most of their mass.  In this respect, there
may be an analogy between protostellar interactions and the interactions
that occur between galaxies, since galaxy interactions are known to play
a major role in driving gas to the centers of galaxies and triggering
starburst and AGN activity there.  AGN activity involves the growth by
accretion of a central massive black hole, and the growth of central
black holes in galaxies could share some similarities with the building
up of massive  stars by tidally induced accretion processes in clusters
(see Section~4.5).

\section{FRAGMENTATION AND TIDAL EFFECTS IN DISKS}

\subsection{The formation of companions}

   If interactions with companions are primarily responsible for
redistributing angular momentum during star formation, the formation
of companions must itself be an intrinsic part of the star formation
process.  The frequency and other statistical properties of binaries
imply that binary or multiple systems are the normal if not universal
outcome of star formation (Larson 2001), and this is also suggested by
the complex structure of many observed star-forming cloud cores (Mundy
et al.\ 2000, 2001).  Moreover, it is possible that even in the more
idealized standard model for isolated star formation, one or more
companion objects may eventually be formed by the fragmentation of
the predicted circumstellar disk.  The disk in this model becomes
gravitationally unstable well before most of the final stellar mass has
been accreted (Stahler 2000), and it can then avoid fragmentation only
if it can settle into a marginally stable state in which spiral density
fluctuations continually transport angular momentum outward.  It is
not clear whether a such marginally stable state can be created and
sustained, as was noted in Section~2.1, but even if it is, the torques
produced by transient spiral density fluctuations may not be strong
enough to drive accretion onto the central star faster than the rate
at which the disk gains mass.  The timescale for transport of angular
momentum by such torques was estimated by Larson (1989) to be of
the order of $10^5$~yr at a radius of 5~au and $10^6$~yr at a radius of
100~au, and this is longer than the predicted timescale for the growth
in mass of the disk (Stahler 2000), especially if the accretion rate
is higher than in the standard model (Larson 1998).  The disk may
then gain mass faster than it can be restructured by internal torques,
making a runaway gravitational instability unavoidable.

   The outcome of such an instability is not yet clear (Durisen
2001), but analytical theory (Shu et al.\ 1990; Adams \& Lin 1993) and
some numerical simulations (Bonnell \& Bate 1994; Burkert, Bate, \&
Bodenheimer 1997; Boffin et al.\ 1998; Nelson et al.\ 1998; Boss 2000;
Bonnell 2001) have suggested that the result is the fragmentation of
the disk and the formation of one or more companion objects.  Other
simulations do not show the formation of bound objects, but only show
some restructuring of the disk by large transient spiral density
fluctuations (Nelson, Benz, \& Ruzmaikina 2000; Pickett et al.\
2000a,b); however, these results are very sensitive to the thermal
behavior of the disk, which has not yet been calculated in full detail.
According to Gammie (2001), if the timescale for radiative cooling
is sufficiently long, a quasi-steady state with transient spiral
fluctuations can exist, while if cooling is sufficiently rapid, the
disk will fragment into clumps.  Recently Boss (2001b) has found, in
simulations that include radiative transfer, that radiative cooling
is effective enough to allow a marginally stable disk to fragment and
form a bound companion object.  The formation of a companion may in
fact be a rather general result, since it represents the lowest energy
state that the matter in the disk can attain subject to conservation
of its angular momentum.  Once a companion has formed, it is no longer
a transient feature of the system, and it can only grow in mass through
further interaction with its surroundings; thus a density fluctuation
large enough to form a bound companion only has to occur once to
change irreversibly the evolution of the system.

   Once formed, a bound companion, as a coherent mass concentration,
will produce stronger and more persistent torques than transient spiral
density fluctuations, and it will drive faster disk evolution.  Even
an object with a mass as small as that of Jupiter could influence the
evolution of a disk by tidally extracting angular momentum from the
inner region and transferring it to the outer region (Goldreich \&
Tremaine 1980, 1982; Lin \& Papaloizou 1993; Goodman \& Rafikov 2001.)
The disturbance created by such an object generates acoustic waves in
the disk that propagate away from the source, and these waves can in
principle transport angular momentum over large distances if they are
not damped out too quickly; for a Jupiter-mass object, this effect
could drive disk evolution on a timescale of $10^6$~years, so it might
be important for the later stages of protostellar accretion (Larson
1989).  Since the tidal torque between a companion object and a disk is
proportional to the square of the companion's mass, objects more massive
than Jupiter will produce much stronger effects, so that even massive
planets could play an important role in redistributing angular momentum
during star formation.  Goodman \& Rafikov (2001) have suggested that
the combined tidal effects of many smaller planets could also provide
an effective viscosity for protostellar disks.

   If the formation of one or more companions is a normal or even
unavoidable part of the star formation process, this may be nature's
way of solving the problem of angular momentum transport.  Since bigger
companions produce stronger effects, the most efficient redistribution
of angular momentum probably occurs in the formation of binary systems
of stars with similar masses; if both stars are surrounded by residual
disks, each star can then extract angular momentum tidally from the
other's disk, causing material to be accreted by both stars, as happens
in numerical simulations of binary formation (Bate \& Bonnell 1997;
Bate 2000).  Most of the mass of a collapsing cloud core can then be
accreted by the two stars and most of its angular momentum can go into
their orbital motions.  Binary systems of stars with similar masses
are, in fact, a common outcome of the star formation process, at least
for relatively close systems, since the typical ratio of secondary to
primary mass in these systems is about 0.5 (Abt 1983; Mayor et al.\
1992, 2001).  Nature therefore seems to have a preference for forming
systems whose properties are optimal for redistributing angular
momentum by tidal torques.

   If tidal interactions play a central role in star formation, star
formation is then clearly a more chaotic and variable process than in
the standard model, and it may not have such a predictable outcome.
Indeed, in this case there may be no `standard model,' and the
processes by which stars form may vary considerably in their details.
One implication of this more chaotic picture of star formation is that
protostellar disks may often have a more violent and irregular early
history than in the standard models, and they may experience strong
disturbances or even disruption as a result of interactions with other
stars.  This clearly has important implications for planet formation,
which will be discussed further in Section~4.2.

\subsection{The effect of tidal perturbations on disks}

   Tidal torques cannot by themselves fully solve the angular momentum
problem, however, since their effects are local and their importance
falls off rapidly with distance from the perturber.  Additional
mechanisms are therefore needed if tidal effects are to influence the
evolution of the inner parts of disks, far from the perturber.  One
possibility is that acoustic waves generated by tidal disturbances
in the outer part of a disk propagate inward and transport angular
momentum as they do so (Spruit 1987, 1991; Larson 1989, 1990a,b; Lin \&
Papaloizou 1993; Lubow \& Artymowicz 2000; Stone et al.\ 2000; Boffin
2001).  A desirable feature of wave transport in this respect is that
any wave with a trailing spiral pattern always transports angular
momentum outward, regardless of the nature or direction of propagation
of the wave (Larson 1989).  Many numerical simulations have shown the
development of trailing spiral wave patterns in tidally disturbed disks,
typically with a two-armed symmetry reflecting the symmetry of the tidal
distortion (Sawada et al.\ 1986, 1987; R\'o\.zyczka \& Spruit 1993;
Savonije, Papaloizou, \& Lin 1994; Murray et al.\ 1999; Haraguchi,
Boffin, \& Matsuda 1999; Bate 2000; Makita, Miyawaki, \& Matsuda 2000;
Blondin 2000).  These tidally generated waves often develop into shocks,
and the associated dissipation of the waves' angular momentum, which
is negative for inward-propagating waves, then reduces the angular
momentum of the disk and drives an inflow (Shu 1976; Spruit et al.\
1987; Spruit 1991; Larson 1989, 1990a; Blondin 2000; Boffin 2001).

   The effect of tidally generated waves on the evolution of a disk
depends on how far the waves can propagate before being damped out,
and this depends on many details of wave propagation and dissipation
that are not yet well understood (Lubow \& Artymowicz 2000).  If
waves can propagate to large distances, they can potentially drive
an accretion flow through the entire region in which they propagate.
For example, if waves produced by tidal disturbances in the outer part
of a protostellar disk can propagate all the way to the region inside
1~au where the magnetorotational instability becomes important, the
combination of tidal waves in the outer region and the magnetorotational
instability in the inner region might be able to drive an inflow through
the entire disk.  Magnetoacoustic waves can also exist if magnetic
coupling is important, and they can contribute to the transfer of
angular momentum.  Understanding better the interaction between waves
and magnetic fields may be an important part of the overall problem of
understanding of the evolution of protostellar disks (Stone et al.\
2000).

   If wave transport does not occur in some regions because the waves
are damped out there (Terquem 2001), inflow may occur only in those
parts of a disk where wave transport is important, and the inflowing
gas may pile up where the waves are damped (Larson 2001).  Local
self-gravity may then eventually become important in these regions,
with the same possible consequences as were discussed above.  Torques
due to transient spiral density fluctuations might restructure the disk
locally, but it is also possible that fragmentation will occur and lead
to the formation of a new companion object such as a small star or
massive planet; disk fragmentation induced by tidal effects has been
found, for example, in the simulations of Boffin et al.\ (1998) and
Horton, Bate, \& Bonnell (2001).  Any new companion formed in this way
could help to drive further disk evolution by its local tidal effects.
The above conjecture that nature solves the angular momentum problem by
forming companions might then be extended to the stronger conjecture
that nature forms as many companions as are needed to redistribute
angular momentum efficiently during the star formation process.

\section{FURTHER IMPLICATIONS OF TIDAL EFFECTS}

\subsection{Early protostellar evolution}
 
   If tidal interactions are important in driving protostellar accretion
at early times, the accretion process will then be time-variable, and
it may occur mostly in bursts triggered by the interactions, as in
the simulations of Bonnell \& Bastien (1992).  Considerable evidence
indicates that the accretion rates of the very youngest stellar objects
do, in fact, vary strongly with time.  The observed luminosities of
the youngest stars are lower than is expected for models with steady
accretion, and this can be understood if most of the accretion occurs in
short bursts (Kenyon \& Hartmann 1995; Calvet, Hartmann, \& Strom 2000).
The jet-like Herbig-Haro outflows, which are believed to be powered
by rapid accretion onto still forming stars, are clearly episodic or
pulsed, and this suggests that the accretion process itself is episodic
(Reipurth 2000).  It is especially noteworthy that the jet sources
frequently have close companions; at least 85 per cent of the jet
sources are members of binary or triple systems, which is the highest
binary frequency yet found among young stars (Reipurth 2000, 2001).
This strongly suggests a causal connection between the presence of a
close companion and the launching of a jet, as would be expected if
tidal interactions were responsible for the episodes of rapid accretion
that produce the jets (Reipurth 2001; Larson 2001).

   The same bursts of rapid accretion that create the jets may also
produce FU~Orionis outbursts, another phenomenon of very early stellar
evolution that is believed to be caused by episodes of rapid accretion
(Dopita 1978; Hartmann \& Kenyon 1996; Reipurth 1989, 2000).  The
possibility that the FU~Orionis phenomenon is caused by tidal
interactions was first suggested by A.~Toomre in 1985, as quoted by
Kenyon, Hartmann, \& Hewitt (1988) and Hartmann \& Kenyon (1996), and it
is supported by the numerical simulations of Bonnell \& Bastien (1992)
which show bursts of rapid accretion triggered by tidal interactions
in a forming binary system.  Bell et al.\ (2000) have suggested that
episodic infall in a complex environment like a forming star cluster
might also play a role in triggering the FU~Orionis phenomenon.  The
most vigorous forms of activity in very young stars might then all
result from bursts of rapid accretion caused by interactions in
systems of forming stars.

\subsection{Planet formation}

   If tidal interactions play an important role in star formation,
this clearly has implications for planet formation too.  Many
planet-forming disks may have had more chaotic and violent early
histories than that postulated in standard `solar nebula' models, and
quiescent disks suitable for forming regular planetary systems like
our own may exist only around relatively isolated stars, many of which
may have been ejected from forming multiple systems.  The properties of
the residual disks around these stars might then be quite variable from
case to case, reflecting their chaotic earlier histories.  Even our Sun
could have been formed in a multiple system in which its protoplanetary
disk was disturbed by interactions; this is suggested by the fact that
the fundamental plane of our planetary system is tilted 8 degrees with
respect to the Sun's equatorial plane, a tilt that could plausibly have
been caused by an encounter with another star in a forming multiple
system (Herbig \& Terndrup 1986; Heller 1993).  Our solar system might
then represent only a particular special case of planet formation, and
it might not be typical.

   A major change in our view of what is typical has, in fact, been
forced by the discovery in recent years of many extrasolar planets that
do not resemble anything seen in our solar system (Marcy \& Butler 1998,
2000; Marcy, Cochran, \& Mayor 2000).  Most of the newly discovered
planets have masses similar to that of Jupiter or larger, yet most
of them have smaller and more eccentric orbits that are not readily
accounted for by the standard models of planet formation.  The broad
distribution of their orbital parameters instead resembles that of
binary systems, and it has been suggested on this basis that they might
have a similar origin (Heacox 1999; Stepinski \& Black 2000, 2001;
Mazeh \& Zucker 2001).  The wide spread in the orbital properties of
both binary systems and extrasolar planets might then result in both
cases from their formation in dense and complex environments where
interactions are frequent (Larson 2001).  For example, if the massive
extrasolar planets were formed in circumstellar disks that were strongly
disturbed by interactions, their formation might have resembled that
of binary companions more than that of the planets in our solar system,
blurring the distinction between star formation and planet formation,
just as the extrasolar planets seem to be blurring the distinction
between stars and planets (Stepinski \& Black 2001).  If they formed
early enough, the known extrasolar planets could have contributed to
the redistribution of angular momentum during the formation of their
associated stars, and their existence would then strengthen the case
for the importance of tidal interactions in star formation.

   Another consequence of a chaotic picture of star formation in which
circumstellar disks are often disturbed by interactions is that shocks
generated by tidal disturbances can produce transient heating events
that might account for the high-temperature inclusions observed in
meteorites.  Chondrules, for example, show evidence for recurrent short
heating events reaching temperatures of the order of 2000~K (Jones et
al.\ 2000).  Standard disk models offer no natural explanation for such
events, but tidally generated shocks could readily produce temperatures
of the required order for the required short times, since this would
need shock speeds of only a few km$\,$s$^{-1}$ which are easily produced
by tidal disturbances.  As discussed by Jones et al.\ (2000), nebular
shocks with Mach numbers of 3 to~8, corresponding to shock speeds of
this order, would provide a viable heating mechanism if there were
a way to generate such shocks.  Tidal disturbances in a dense
star-forming environment might provide the required mechanism.  In
this case the chondrules may bear witness to a violent early history
of our solar system.

\subsection{Formation of massive stars}

   Massive stars must evidently accrete more mass when they form
than low-mass stars, and more angular momentum must therefore be
redistributed during this process by mechanisms such as tidal torques.
Because radiation pressure and winds from massive stars can blow away
diffuse gas, the matter accreted by a forming massive star must also be
very dense and probably very clumpy to overcome these effects (Larson
1982, 1999; Bonnell, Bate, \& Zinnecker 1998; Stahler, Palla, \& Ho
2000).  Massive stars may therefore only be able to form in very dense
and complex environments like those found in dense forming clusters,
where interactions with other nearby stars and dense clumps can
contribute to the accretion process.  Massive stars do, in fact, tend to
form in very dense environments, and preferentially near the centers of
dense clusters (Larson 1982; Zinnecker, McCaughrean, \& Wilking 1993;
Hillenbrand \& Hartmann 1998; Garay \& Lizano 1999; Clarke, Bonnell, \&
Hillenbrand 2000; Stahler et al.\ 2000).  Since stronger torques are
needed to transfer angular momentum away from the more massive forming
stars, more numerous and/or more massive close companions may also be
required than is the case for the less massive stars; this possibility
is consistent with the fact that massive stars have a particularly high
frequency of close companions, and often occur in multiple systems with
other massive stars (Mason et al.\ 1998; Preibisch et al.\ 1999; Stahler
et al.\ 2000; Preibisch, Weigelt, \& Zinnecker 2001; Mermilliod \&
Garc\'\i a 2001).

   If the companions that help to redistribute angular momentum during
the formation of the massive stars were themselves formed by processes
involving interactions with less massive stars, this suggests that
massive stars are formed by a bootstrap process that builds up stars
of progressively larger mass in a hierarchical fashion (Larson 1999).
The most massive stars may then represent the culmination of such a
process, forming as a result of a series of many interactions in a very
dense and complex environment.  Such a chaotic and complex formation
process contrasts with that envisioned in models which postulate a
smooth and centrally condensed structure for massive collapsing cloud
cores (Garay \& Lizano 1999).  At present, the formation of massive
stars remains very poorly understood, both observationally and
theoretically, and the best evidence we have concerning how they form
may be the fossil evidence provided by systems of young massive stars.
The high incidence of close companions to these stars, and the fact
that these companions are themselves relatively massive, suggest that
gravitational interactions have played a particularly important role
in the formation of the most massive stars.  Such interactions might
even include direct collisions and mergers between newly formed stars
(Bonnell et al.\ 1998) or star-forming cores (Stahler et al.\ 2000).
Mergers might be regarded as an extreme type of accretion event, which
as before requires angular momentum to be lost from the material
being accreted; in this case the angular momentum may be lost through
interactions with other forming stars and gas in a dense environment,
perhaps similar to the processes involved in the formation of close
binaries (Section~2.2).  The building up of massive stars by mergers
might then be viewed as an extension of the formation of close binaries
to the case where so much angular momentum is lost that a merger occurs.

\subsection{The stellar IMF}

   If massive stars are built up progressively by interactions with
less massive stars, this has implications for the relative numbers
of stars of different masses that can be formed.  The tidal effects
discussed above imply a dynamical coupling between the masses of stars
that form near each other, since the most efficient redistribution of
angular momentum occurs when stars interact with other stars of similar
or not much smaller mass.  As a result, a star of any given mass is
most likely to form in association with other stars that have similar
masses.  As was noted above, the stars in close binary systems tend to
have masses that are more nearly equal than would be expected if they
had been randomly selected from a standard IMF (Abt 1983; Mayor et al.\
1992, 2001), showing that the masses of stars that form near each other
are correlated (Larson 2001).  There should then be some relation
between the numbers of stars that form in adjacent mass intervals,
and the resulting IMF should be a continuous function without large
gaps in mass; this would avoid the problem of simple accretion models
(Zinnecker 1982) in which a large mass gap is created by the runaway
growth in mass of the largest object.  If the numbers of stars in
adjacent mass intervals are coupled by dynamical effects that depend
only the mass ratio and not on the absolute mass, the resulting IMF
will then be a scale-free power law, consistent with the evidence that
the upper IMF has a power-law form similar to the original Salpeter
function (Scalo 1998; Larson 1999; Meyer et al.\ 2000).

   In order to redistribute angular momentum effectively during the
formation of the more massive stars, the less massive ones probably
must contain a comparable or larger total amount of mass.  If, for
example, efficient redistribution of angular momentum through the mass
hierarchy requires the same amount of mass in each logarithmic mass
interval, this would correspond to an IMF with a logarithmic slope of
$x = 1$ which is not very different from the Salpeter slope $x = 1.35$.
With the Salpeter slope, the amount of mass per unit logarithmic mass
interval increases by about a factor of 2 with each factor of 10
decrease in stellar mass, making the less massive stars moderately
dominant in mass, and plausibly allowing them to redistribute angular
momentum effectively enough to enable the formation of the more
massive stars.

   While these considerations are qualitative and do not yet provide
any quantitative prediction of the slope of the upper IMF, this view of
the origin of the IMF differs fundamentally from most previous theories
in that gravitational dynamics here plays a central role in determining
stellar masses and the IMF.  The building up of the upper IMF then
occurs by processes that are at least partly deterministic in nature,
in contrast with most previous theories where the origin of the IMF is
based on statistical or geometrical hypotheses (Larson 1999).  If the
form of the upper IMF results from universal processes of gravitational
dynamics rather than from complex astrophysical processes, this would
make it easier to understand why the slope of the upper IMF seems to be
universal and shows no clear dependence on any astrophysical parameters.

\subsection{The growth of massive black holes}

   It is also possible that there are similarities between the star
formation processes discussed here and the growth of black holes in
galactic nuclei.  In both situations, viscous accretion disks have been
invoked to transfer angular momentum outward and mass inward, and
similar problems have been encountered, such as a tendency for these
disks to become gravitationally unstable (Shlosman \& Begelman 1989;
Begelman 1994; Larson 1994; Menou \& Quataert 2001).  As is the case for
star formation, it is important to understand the fate of the angular
momentum extracted from the matter that goes into a nuclear black hole,
and it is difficult to see how all of this angular momentum could go
into an extended gas disk; more plausibly, most of it goes into the
orbital motions of other condensed objects such as stars or other black
holes.  Again, this requires that tidal torques remove the excess
angular momentum from the gas being accreted, and these torques
must come from large mass concentrations.  One possibility is that
interactions between two massive black holes in a merging system of
galaxies may play an important role.  Another possibility is that
gravitational instability in the dense gas orbiting around a growing
central black hole produces new mass concentrations such as massive
clusters (Heller \& Shlosman 1994).  Collective modes such as bars may
also play a role on larger scales.  In any case gravitational effects
of some kind seem likely to be involved in the black hole accretion
process (Larson 1994).

   On larger scales, galaxy interactions and mergers clearly play
important roles in redistributing angular momentum and feeding gas into
the nuclear regions of galaxies, and they provide a clear example of
tidal torques at work.  Many of the phenomena that have been discussed
in this paper may thus be small-scale analogs of processes that also
occur on galactic scales, where they have been better studied because
observations are easier.  The formation of massive stars in clusters
and the formation of massive black holes in galaxies may be closely
related problems.

\section{SUMMARY AND CONCLUSIONS}

   Although magnetic fields may remove much of the angular momentum
from star-forming clouds during early stages of their contraction, this
still leaves an important part of the angular momentum problem unsolved
because these fields decouple from the gas long before stellar densities
are reached.  Angular momentum is then approximately conserved during
the later stages of the collapse, and the angular momentum observed in
collapsing cloud cores is still three orders of magnitude larger than
the maximum that can be contained in a single star.  Nearly all of the
angular momentum of the matter that condenses into each forming star
must therefore be removed or redistributed during the formation process,
and it could in principle go either into diffuse gas, for example in an
extended disk as in standard accretion disk models, or into the orbital
motions of the companion stars in a binary or multiple system.

   Internal transport processes in disks are probably not sufficient
by themselves to solve the angular momentum problem, since even if one
or more such mechanism does operate, the transport timescale becomes
unacceptably long if the disk expands to the very large radius required
by standard disk models.  However, tidal interactions with other stars
in a forming binary or multiple system can plausibly be of very general
importance in redistributing angular momentum during the star formation
process.  Statistical evidence suggests that most, if not all, stars
do indeed form in such systems, and in this case most of the angular
momentum of a collapsing cloud core can go into the stellar orbital
motions.  The tidal torques acting in such systems can then effectively
transfer angular momentum from the gas orbiting around each forming
star to the orbital motions of the companion stars.

   If tidal interactions with companions remove most of the excess
angular momentum from forming stars, the formation of companions must
itself be an intrinsic part of the star formation process.  Observed
star-forming cloud cores often show internal structure that is likely
to form binary or multiple systems, but even in a more idealized
situation such as the standard model for isolated star formation,
one or more companions may eventually form by the fragmentation of a
circumstellar disk.  Companions formed in this way might include massive
planets, which can also play a role in redistributing angular momentum
by their tidal effects on the remaining disk.  Waves generated in disks
by tidal disturbances may contribute to the overall transport of angular
momentum in the system.  If the formation of companions is indeed a
normal or even unavoidable part of the star formation process, this
may then be nature's way of solving the problem of angular momentum
transport.

   If tidal interactions are responsible for redistributing angular
momentum and driving protostellar accretion, the accretion process
must be time-variable, and it may occur mostly in bursts triggered by
the interactions.  Tidally induced bursts of rapid accretion may account
for some of the more energetic activity of very young stars, such as
the jet-like Herbig-Haro outflows; a connection between interactions
and outflows is indeed suggested by the fact that the outflow sources
have an unusually high frequency of close companions (Reipurth 2000).
The early evolution of planet-forming disks may also often be a less
regular and more chaotic process than in standard models, and tidally
induced shock heating events might account for the heating of the
chondrules in meteorites.

   Interactions with companions are probably especially important
for the formation of massive stars, since these stars form in dense
environments and have a high frequency of close companions which are
themselves relatively massive.  If stars gain most of their mass through
interactions with companions that have similar or not much smaller
masses, the numbers of stars that form in adjacent mass intervals will
then be dynamically coupled, and this will have important implications
for the form of the stellar IMF.  If the form of the upper IMF is
determined by universal processes of gravitational dynamics that have
no intrinsic mass scale, this could help to explain why the upper IMF
appears to have a nearly universal power-law form and shows no clear
dependence on any astrophysical parameters.

   Many of the processes that have been discussed here have analogs on
galactic scales, where tidal interactions clearly play a very important
role in driving gas to the centers of galaxies and triggering starburst
and AGN activity there.  The growth of massive black holes in galactic
nuclei may involve processes similar to those involved in the formation
of massive stars in clusters, and in this case the formation of massive
black holes in galaxies and the formation of massive stars in clusters
may be closely related problems.

\section*{Acknowledgments}

   Some of the ideas elaborated in this paper emerged from discussions
at IAU Symposium 200 on The Formation of Binary Stars held in Potsdam,
April 2000, and I acknowledge particularly the contributions of Hans
Zinnecker, Robert Mathieu, and Bo Reipurth to the organization of the
meeting and to the scientific discussions.  I also acknowledge useful
comments from the referee that led to a number of significant
improvements in this paper.

\label{lastpage}

\end{document}